\documentclass[english]{article}

\usepackage[T1]{fontenc}
\usepackage[latin9]{inputenc}
\usepackage{amsmath}
\usepackage{babel}
\usepackage{color}

\definecolor{lightgray}{rgb}{.7,.7,.7}

\definecolor{red}{rgb}{1,0,0}
\def\red{\color{red}}
\definecolor{green}{rgb}{0,1,0}

\definecolor{blue}{rgb}{0,0,1}

\newtheorem{definition}{Definition}

\newcommand{\hR}{{\hat R}}
\newcommand{\heq}{{\ {\hat =}\ }}
\newcommand{\cL}{{\cal L}}
\newcommand{\cH}{{\cal H}}
\newcommand{\hvar}{{\hat\varepsilon}}
\newcommand{\he}{{\hat e}}
\newcommand{\hg}{{\hat g}}
\newcommand{\tna}{{\tilde{\nabla}}}

\newcommand{\tR}{{\tilde{R}}}

\newcommand{\bR}{{\bar R}}

\newcommand{\bGa}{{\bar\Gamma}}
\newcommand{\tGa}{\tilde{\Gamma}}
\newcommand{\hth}{{\hat\theta}}

\begin{document}
\begin{center}
{\bf\Large Fluid/Gravity Correspondence For General Non-rotating
Black Holes}
\end{center}
\begin{center}
Xiaoning Wu${}^{1,2,3}$, Yi Ling${}^{5,3}$, Yu Tian${}^{4,3}$ and Chengyong Zhang${}^4$\\
1. Institute of Mathematics, Academy of Mathematics and System Science, Chinese Academy of Sciences, Beijing 100190, China\\
2. Hua Loo-Keng Key Laboratory of Mathematics, {CAS}, Beijing
100190, China\\{3. State Key Laboratory of Theoretical Physics,
Institute of Theoretical Physics, Chinese Academy of Sciences,
Beijing 100190}\\ 4. School of Physics, University of
Chinese Academy of Sciences, Beijing 100049, China\\
5. Institute of High Energy Physics, Chinese Academy of Sciences,
Beijing 100190, China
\end{center}
\begin{abstract}
In this paper, we {investigate} the fluid/gravity correspondence {in
spacetime with} general non-rotating weakly isolated horizon. With
the help of Petrov-like boundary condition and large mean curvature
limit, we show that the dual {hydrodynamical} system is {described
by} a generalized forced incompressible Navier-Stocks equation.
{Specially}, for {stationary black holes or those spacetime with
some asymptotically stationary conditions}, such {a} system
{reduces} to a standard forced Navier-Stocks system.
\end{abstract}

\section{Introduction}
The correspondence between gravity and fluid was found in 70's of
last century. Damour\cite{D70} {firstly} noticed that there was a
formal correspondence between the Raychaudury equations and fluid
equations. He suggested that the black hole horizon could be viewed
as a membrane of fluid. Such idea had been studied by other
researchers\cite{SGE} and this method was called "membrane paradigm
of black hole"\cite{PTM86}. These work imply that there should be
some relations between the gravitational perturbation of black hole
horizon and the dynamics of fluid membrane.

More than twenty years later, such a topic has been reconsidered in
the framework of AdS/CFT correspondence. Based on AdS/CFT
correspondence, the gravity in bulk should be dual to some quantum
field theory on the boundary of spacetime. Under certain conditions,
any quantum field theory can be effectively described by
hydrodynamics {at} the long wavelength limit. So there should be a
natural correspondence between the dynamics of long wavelength
gravitational perturbation and hydrodynamics. Such correspondence
was {firstly} studied by Policastro, Son and Starinets \cite{PSS01}.
They considered the hydrodynamics of super-symmetric gauge theory on
the conformal boundary of spacetime and found the similar properties
as found by Damour on black-hole horizon. Following this idea, many
important works have been done and there is a nice review on this
topic\cite{SS07}. {This so-called fluid/gravity duality were further
deepened by Bhattacharyya et al to a full correspondence involving
the nonlinear fluid dynamics\cite{BHMR}.} The similarity between
{the fluid/gravity duality in the AdS/CFT framework and black-hole
membrane indicates that there should be some relations between them.
{Indeed, this kind of universality is considered in \cite{IL}, and
is later interpreted as the Wilson renormalization group flow in the
AdS/CFT framework\cite{BKLS}. Some further developments following
this approach can be found in \cite{BKLS11,Cai,NTWL}.}}

The membrane paradigm has been generalized to some asymptotic AdS
black holes as well. Kovtun, Son and Starinets first found that
long-time, long-distance fluctuations of plane-symmetric horizons
exhibit universal hydrodynamic behavior\cite{KSS03}. Later, Eling,
Oz and their colleagues (EO) found\cite{Oz09}, with the
non-relativistic limit and long wavelength limit, that the
gravitational perturbation of black brane {satisfies the}
Navier-Stocks equation. EO method is a quite nice way to establish
the membrane paradigm of horizon because it is a local method which
only requires the geometric information near horizon but do not need
the knowledge of the asymptotic region. But this approach contains
some drawbacks. As emphasized by them, the long wavelength limit
plays a crucial role in this method, thus it can only be applicable
to the spatially non-compacted case, typically in which the horizon
is plane-symmetric. Some improvement has been made recently. In
Ref.\cite{LS-PE2NS}, Lysov and Strominger(LS) studied the
fluid/gravity correspondence with the help of Petrov-like boundary
condition in a flat spacetime. They found that, with the help of
Petrov-like boundary condition and large mean curvature limit (which
is equivalent to near horizon limit), the correspondence between
Einstein equation and Navier-Stocks equation can be established. The
main idea of their work is following. First, they consider the
perturbations of the extrinsic curvature $K_{ab}$ of a hyper-surface
$\Sigma_c$ and impose a Petrov-like boundary condition for Weyl
curvature on $\Sigma_c$. Since the Weyl tensor is traceless, the
Petrov-like boundary condition provides $\frac{p(p+1)}{2}-1$
constraints on the $\frac{(p+1)(p+2)}{2}$ components of $K_{ab}$.
The Gaussian equation also {gives} another constrain equation for
the perturbation of $K_{ab}$, so the remaining $p+1$ independent
components of $K_{ab}$ may be interpreted as {the} velocity field
$v^{i}$ and the pressure $P$ of fluid living on this hypersurface.
The Gaussian equation on $\Sigma_{c}$ can be viewed as an equation
of state for this fluid and the Codazzi equations on $\Sigma_{c}$
will give the evolution equation of this fluid. Then taking a
suitable non-relativistic limit and large mean curvature limit (near
horizon limit), the Codazzi equations have the form of
incompressible Navier-Stocks equations. In Ref.\cite{BKLS11},
Strominger and his colleague have shown that the long wave
perturbation solution of gravity satisfies the Petrov-like boundary
condition. This implies that we can use the Petrov-like boundary
condition instead of long-wave condition to evade the difficulty of
compactness of the hypersurface. Like EO's method, LS's method is
also a local method in linking the Einstein equation and the
Navier-Stocks equation. Since in this approach one only imposes
Petrov type boundary condition and the near horizon limit rather
than the long wavelength limit, the requirement of the spatial
uncompactness for the hypersurface becomes unnecessary. Some
generalizations of Strominger's work have been considered.
Ref.\cite{HLPTW_PE2NS-curvedspatial,HLPTW_PE2NS-cosmol} generalized
this framework to some curved cases and the cases with a
cosmological constant, respectively. Ref.\cite{ZLNTW_MG} further
considered the situation of spacetime with electromagnetic field.

It is worthwhile to point out that, up to now the fluid/gravity
correspondence in this route has only been investigated in some
concrete background such as static black hole solutions in
literature. However, based on the accumulated experience in the
study of AdS/CFT, one has reasons to believe that such a
correspondence should be a quite general notion. At least, it should
be applicable to general stationary black holes, so a general proof
for such correspondence {in a more general setting } is needed. This
paper will focus on this topic. We will show that for any spacetime
which contains a non-rotating weakly isolated horizon LS's
realization of the fluid/gravity correspondence {can} always be
established.

This paper is organized as following{.} In Sec.II, the definition
of weakly isolated horizon is introduced and the near horizon
geometry is studied. In Sec.III, we introduce the Petrov-like
condition for gravity and {analyze} its detailed behavior under
the near horizon limit. The equivalence between Strominger's large
mean curvature limit and near horizon limit {has been addressed
and the} Gaussian equation {has also} been considered in this
section. In Sec.IV, we derive the incompressible Navier-Stokes
equation from the Codazzi equation for gravity. Section V contains
some discussions of our work.

\section{The geometry near non-rotating isolated horizon}
In order to consider the existence of fluid/gravity correspondence
for general black hole, a general definition of black hole horizon
is needed. {The weakly isolated horizon (WIH) advocated by Ashtekar
{\it et. al.}} is a nice choice\cite{AK04}. This {sort of horizon
preserves} many important properties of traditional black hole
horizon{, but be applicable to} more general cases. It has been
shown that stationary black hole horizon are all
WIH\cite{AK04,KLP05}. Roughly speaking, WIH is a {non-expansion}
light cone with almost stationary inner geometry. A nice review of
WIH can be found in \cite{AK04}. In ref.\cite{KLP05}, Ashtekar's
definition has been generalized to higher dimensional space-time.
The definition of WIH in $(p+2)$-dimensional space-time is {given}
as following\cite{KLP05}.

\begin{definition}
(Weakly Isolated Horizon in $(p+2)$-dimensional space-time)\\
Let $(M,g)$ be a $(p+2)$-dim Einstein manifold with or without {a}
cosmological constant{\red.} $\cH$ is a $(p+1)$-dim null
hypersurface in $M$ {and} $l$ is the null normal of $\cH$. $\cH$ is called a weakly isolated horizon in $M$ if\\
(1). there exists an embedding : $S\times [0,1]\to M$, $\cH$ is the
image of this map, $S$ is a p-dimensional compact, connected
manifold and for every maximal null curve in $\cH$ there exists
$x\in S$ such that the curve is the image
of $x\times[0,1]$;\\
(2). the expansion of $l$ vanishes
everywhere on $\cH$;\\
(3). $R_{ab}l^al^b|_{\cH}=0$;\\
(4). let ${\cal D}$ denote the induced connection on $\cH$, $[\cL_l,
{\cal D}]l=0$ holds on $\cH$.
\end{definition}

To generalize fluid/gravity correspondence to WIH case, the geometry
near horizon is needed. So in this section, we will get some control
on the behavior of metric near horizon.

Suppose ${\cal H}$ is a WIH in a $(p+2)$-dimensional spacetime,
based on the definition of ref.\cite{AK04,KLP05}. To control
geometry near a null hypersurface, a Bondi-like coordinate {system
is} always a convenient choice\cite{Fr81}. Let $l$ be the tangent
vector of null generator of ${\cal H}$ with parameter $t$. The level
set of $t$ in ${\cal H}$ gives a space-like foliation $\{S_t\}$ of
${\cal H}$. On section $t=0$, there {are} coordinates $\{x^i\}$.
Generator of horizon will bring this coordinates to the whole
horizon. On each section $S_t$, let $\{E_I\}$ be a set of
orthonormal basis for the tangent space of $S_t$. Under some
suitable rotation, $\{E_I\}$ can always be chosen such that
$(\cL_lE_I)|_{\cal H}\propto E_I$. Therefore, we have tetrad
$(l,E_I)$ at each point of ${\cal H}$. Then choosing past-pointed
null vector field $n$ on ${\cal H}$ such that $\langle l,n\rangle=1$
and $\langle n, E_I\rangle=0$. It is easy to see that $n$ is unique
at each point of $\cH$.  For any $p\in {\cal H}$, there {exists a}
unique null geodesic $\gamma$ with respect to $n$. The affine
parameter of $\gamma$ is $r$ and $\gamma(0)=p$. One can extend
coordinates $(t, x^i)$ into spacetime by Lie dragging them along
$\gamma$ and $(t,r,x^i)$ is a coordinates system near horizon.
Similarly, one can also extend the null tetrad $(l,n,E_I)$ into the
spacetime by requiring $\nabla_n(l,n,E_I)=0$. We call coordinates
$(t,r,x^i)$ as a Bondi-like coordinates near horizon $\cal H$ and
above chosen tetrad as Bondi-like tetrad near horizon. With this
choice of Bondi-like coordinates, the Bondi-like tetrad can be
expressed as
\begin{eqnarray}
n & = & \partial_{r},\nonumber \\
l & = & \partial_{t}+U\partial_{r}+X^{i}\partial_{i},\nonumber\\
E_{I} & = & W_{I}\partial_{r}+e_{I}^{i}\partial_{i},\qquad I,i=1,2,\cdots,p,\label{tetrad0}
\end{eqnarray}
where $(U,X^{i},W_{I},e_{I}^{i})$ are functions of $(t,r,x^i)$ and
satisfy $U\heq X^i\heq W_I\heq 0$. (Note: Here we follow the
notation of ref.\cite{AK04}, ``$\heq$'' means equality holds only on
horizon $\cal H$. In the following of this paper, we also use $\hat
f$ to denote the value of function $f$ on horizon.) Using the null
tetrad, the metric can be expressed as
$g^{ab}=l^an^b+n^al^b+E^a_IE^b_I$, then components of metric are :
\begin{equation}
(g^{\mu\nu})=\left(\begin{array}{ccc}
 0 & 1 & \vec{0}\\
 1 & 2U+W_{I}W_{I} & X^{i}+W_{I}e_{I}^{i}\\
 \vec{0} & X^{j}+W_{I}e_{I}^{j} &e_{I}^{i}e_{I}^{j}
\end{array}\right).\label{eq:metric}
\end{equation}
With this Bondi gauge, it is also easy to see that following relation holds:
\begin{eqnarray}
&&\alpha_{I}:=-\left\langle l,\nabla_In\right\rangle,\quad \pi_{I}:=\left\langle E_{I},\nabla_{l}n\right\rangle\nonumber\\
&&\alpha_I+\pi_I=0
\end{eqnarray}
Based on the discussion of ref.\cite{AK04,KLP05}, $\pi_I$ is related
with the angular momentum of horizon. In this paper, we consider
non-rotating black holes, so we require the non-rotating condition :
$\pi_I\heq 0$.

In order to get the behavior of metric near horizon, let's consider the Cartan structure equations,
\begin{eqnarray}
\left[n,E_{I}\right]
& = & \frac{\partial W_{I}}{\partial r}\partial_{r}+\frac{\partial e_{I}^{i}}{\partial r}\partial_{i}\nonumber\\
 & = & \alpha_{I}\partial_{r}-\theta'_{JI}\left(W_{J}\partial_{r}+e_{J}^{i}\partial_{i}\right),\nonumber\\
 \left[n,l\right]
& = & \frac{\partial U}{\partial r}\partial_{r}+\frac{\partial X^{i}}{\partial r}\partial_{i}\nonumber\\
 & = & \varepsilon\partial_{r}-\pi_{I}\left(W_{I}\partial_{r}+e_{I}^{i}\partial_{i}\right).
\end{eqnarray}
Above equations imply
\begin{eqnarray}
\frac{\partial W_{I}}{\partial r}=\alpha_{I}-\theta'_{JI}W_{J},\quad
\frac{\partial e_{I}^{i}}{\partial r}=-\theta'_{JI}e_{J}^{i},\quad
\frac{\partial U}{\partial r}=\varepsilon-\pi_{I}W_{I},\quad
\frac{\partial X^{i}}{\partial r}=-\pi_{I}e_{I}^{i},\label{1o}
\end{eqnarray}
 where $\theta'_{JI}:=\left\langle E_{J},\nabla_In\right\rangle,\varepsilon:=\left\langle n,\nabla_{l}l\right\rangle$.
 Based on the discussion of ref.\cite{AK04,KLP05}, $\varepsilon|_{\cH}$ is a constant and just the surface gravity of
 horizon $\cH$. Combining with Eq.(\ref{eq:metric}), we get the first order derivative of metric.

To control the order of the metric more accurately, the second derivation
of metric are needed. Taking r-derivative on both sides of equations in (\ref{1o}),
\begin{eqnarray}
\frac{\partial^{2}U}{\partial r^{2}} & = & \frac{\partial\varepsilon}{\partial r}-\frac{\partial\pi_{I}}{\partial r}W_{I}-\pi_{I}\frac{\partial W_{I}}{\partial r},\nonumber \\
\frac{\partial^{2}W_{I}}{\partial r^{2}} & = & \frac{\partial\alpha_{I}}{\partial r}-\frac{\partial\theta'_{JI}}{\partial r}W_{J}-\theta'_{JI}\frac{\partial W_{I}}{\partial r},\label{eq:secondderiv}\\
\frac{\partial^{2}X^{i}}{\partial r^{2}} & = & -\frac{\partial\pi_{I}}{\partial r}e_{I}^{i}-\frac{\partial e_{I}^{i}}{\partial r}\pi_{I},\nonumber
\end{eqnarray}
so the values of $\partial_{r}\varepsilon,\partial_{r}\pi_{I},\partial_{r}\alpha_{I}$
and $\partial_{r}\theta'_{JI}$ are needed. With the help of second Cartan structure equations,
\begin{eqnarray}
\partial_r\varepsilon = \nabla_{n}\varepsilon=\nabla_{n}\left\langle n,\nabla_{l}l\right\rangle = R_{nlnl}-\alpha_{I}\pi_{I}.
\end{eqnarray}
Similarly, we have
\begin{eqnarray}
\partial_r\pi_{I} & = & R_{nlIn}-\theta'_{IJ}\pi_{J},\nonumber\\
\partial_r\alpha_{I} & = & R_{nInl}-\alpha_{J}\theta'_{JI},\nonumber\\
\partial_r\theta'_{JI} & = & R_{nIJn}-\theta'_{JK}\theta'_{KI}.
\end{eqnarray}
Plugging these equations in to (\ref{eq:secondderiv}), we have
\begin{eqnarray}
\frac{\partial^{2}U}{\partial r^{2}} & = & R_{nlnl}-R_{nlIn}W_{I}-2\alpha_{I}\pi_{I}+2\theta'_{IJ}\pi_{J}W_{I},\nonumber\\
\frac{\partial^{2}W_{I}}{\partial r^{2}} & = & R_{nInl}-R_{nIJn}W_{J}-\alpha_{J}\left(\theta'_{JI}+\theta'_{IJ}\right)+W_{J}\theta'_{JK}\left(\theta'_{KI}+\theta'_{IK}\right),\nonumber\\
\frac{\partial^{2}X^{i}}{\partial r^{2}} & = & -R_{nlIn}e_{I}^{i}+2\theta'_{IJ}\pi_{J}e_{I}^{i}.
\end{eqnarray}
Using non-rotating condition $\pi_{I}\heq 0$, all unknown functions near horizon are
\begin{eqnarray}
U & = & \hvar r+\frac{1}{2}\hR_{nlnl}r^{2}+O\left(r^{3}\right),\nonumber\\
W_{I} & = & \frac{1}{2}\hR_{nInl}r^{2}+O\left(r^{3}\right),\nonumber\\
X^{i} & = & \frac{1}{2}\hR_{nInl}\he_{I}^{i}r^{2}+O\left(r^{3}\right),\nonumber\\
e^i_J &=& \he^i_J-{\hat\theta}'_{IJ}\he^i_Ir+O(r^2).\label{tetrad}
\end{eqnarray}
Finally, the asymptotic extension of metric near horizon are
\begin{eqnarray}
g^{tr}&=&1,\qquad g^{ti}\ =\ 0,\nonumber\\
g^{rr} & = & 2U+\sum_I W_{I}^{2}=2\hvar r+\hR_{nlnl}r^{2}+O\left(r^{3}\right),\nonumber\\
g^{ri} & = & X^{i}+\sum_I W_{I}e_{I}^{i}=\hR_{nInl}e_{I}^{i}r^{2}+O\left(r^{3}\right),\nonumber\\
g^{ij} & = & \sum_I e_{I}^{i}e_{I}^{j}\sim O\left(r^{0}\right).\label{eq:-gij}
\end{eqnarray}
and
\begin{eqnarray}
g_{tt} & = & -g^{rr}+g^{ri}g_{ij}g^{rj}=-2\hvar r-\hR_{nlnl}r^{2}+O\left(r^{3}\right),\nonumber\\
g_{ti} & = & -g_{ij}g^{rj}=-\hR_{nInl}\hg_{ij}\he_{I}^{j}r^{2}+O\left(r^{3}\right),\nonumber\\
g_{tr}&=&1,\qquad g_{ri}\ =\ 0, \qquad g_{ij}\sim O(r^0).\label{gij}
\end{eqnarray}

\section{The Petrov-like condition for non-rotating weakly isolated horizon}

{As firstly pointed out by Strominger et.al.,  the Petrov-like
boundary condition plays an essential role in the construction of
the gravity/fluid correspondence. The main reason is that imposing
such a condition can guarantee that no out-going gravitational
radiation across the boundary}\cite{LS-PE2NS}. Originally, the
Petrov condition is proposed to classify the geometry of the whole
spacetime. But here {such conditions are only specified on the
cutoff surface, thus one may call it Petrov-like boundary
condition}.

With previous result, for a cutoff surface $\Sigma_{c}:=\{p\in
M|r(p)=r_c\}$ , the induced line element on $\Sigma_c$ could be
written as
\begin{equation}
ds_{p+1}^{2}=g_{tt}dt^{2}+2g_{ti}dtdx^{i}+g_{ij}dx^{i}dx^{j},\ (i,j=1,...,p).
\end{equation}
The Petrov-like condition for gravity is defined by \cite{LS-PE2NS}
\begin{equation}
C_{\mu\nu\rho\sigma}l^{\mu}E_{i}^{\nu}l^{\rho}E_{j}^{\sigma}|_{\Sigma_c}=C_{lilj}|_{\Sigma_c}=0,\label{eq:Petrov0}
\end{equation}
where $C_{\mu\nu\rho\sigma}$ is the Weyl tensor of space-time,
$l^{\mu}$ and $E^{\mu}$ are the null tetrad {as introduced } in {the
} previous section.

It has been shown in the last section that $l^{\mu}$ has the form
$l=U\partial_{r}+\partial_{t}+X^{i}\partial_{i}$. From the normal
covector $N_{a}=(dr)_{a}/\sqrt{g^{rr}}$ of $\Sigma_{c}$, one
obtains
$$\left(\partial_{r}\right)^{a}=\frac{1}{\sqrt{g^{rr}}}N^{a}-\frac{1}{g^{rr}}\left(\partial_{t}\right)^{a}-\frac{g^{ri}}{g^{rr}}\left(\partial_{i}\right)^{a}.$$
Thus

\begin{eqnarray}
l^{a} & = & \frac{U}{\sqrt{g^{rr}}}N^{a}+\left(1-\frac{U}{g^{rr}}\right)\left(\partial_{t}\right)^{a}+\left(X^{i}-\frac{Ug^{ri}}{g^{rr}}\right)\left(\partial_{i}\right)^{a}.
\end{eqnarray}
The Petrov-like condition could be expressed as

\begin{eqnarray}
0 & = & \left[\frac{U^{2}}{g^{rr}}C_{NiNj}+\left(1-\frac{U}{g^{rr}}\right)^{2}C_{titj}+\frac{U}{\sqrt{g^{rr}}}\left(1-\frac{U}{g^{rr}}\right)\left(C_{Nitj}+C_{Njti}\right)\right.\nonumber\\
 &  & +\frac{U}{\sqrt{g^{rr}}}\left(X^{k}-\frac{Ug^{rk}}{g^{rr}}\right)\left(C_{Nikj}+C_{Njki}\right)\nonumber \\
 &  & +\left(X^{k}-\frac{Ug^{rk}}{g^{rr}}\right)\left(X^{m}-\frac{Ug^{rm}}{g^{rr}}\right)C_{kimj}\nonumber \\
 &  & \left.+\left(1-\frac{U}{g^{rr}}\right)\left(X^{k}-\frac{Ug^{rk}}{g^{rr}}\right)\left(C_{tikj}+C_{tjki}\right)\right]_{\Sigma_c}.\label{eq:PetrovExp}
\end{eqnarray}

It has been mentioned in {the section of} introduction, the
intrinsic metric of the cutoff surface is fixed, only the extrinsic
curvature is perturbed. So the {relation} between the spacetime Weyl
tensor and the curvature of the cutoff surface is needed. In the
absence of matter {field}, the spacetime Weyl tensor may be
decomposed by the curvature of the cutoff surface.
\begin{eqnarray}
C_{abcd} & = & \bar{R}_{abcd}-K_{ac}K_{bd}+K_{ad}K_{bc},\nonumber \\
C_{abcN} & = & D_{a}K_{bc}-D_{b}K_{ac},\label{eq:Cdecomp}\\
C_{aNbN} & = & -\bar{R}_{ab}+KK_{ab}-K_{ac}K_{b}^{c}.\nonumber
\end{eqnarray}
Here $D$ is the induced connection on $\Sigma_c$ and
$\bar{R}_{abcd}$ is the associated Riemann curvature. Thus
Eq.(\ref{eq:PetrovExp}) can be expressed in terms of the intrinsic
and extrinsic curvature of the cutoff surface.

{In} the original method {proposed by} Strominger\cite{LS-PE2NS}, to
get the fluid/gravity correspondence, one should consider the near
horizon limit and non-relativistic limit. Such two {limits} can be
{implemented} as following: introduce a rescalling parameter
$\lambda$, define new time coordinate $\tau=2\hvar\lambda^{2}t$ and
choose $r_c=2\hvar\lambda^2$ , consider the limit $\lambda\to 0$,
where $\hvar=\varepsilon|_{{\cal H}}$ is the surface gravity which
has been introduced in last section. In the following parts, we will
consider the Petrov-like condition in such {limits}.

As a result of Eq.(\ref{tetrad}),
\begin{eqnarray}
X^{k}-\frac{Ug^{rk}}{g^{rr}} & = & X^{k}-\frac{UX^{k}+UW_{I}e_{I}^{k}}{2U+W_{I}W_{I}}\sim O\left(r^{3}\right),
\end{eqnarray}
Eq.(\ref{eq:PetrovExp}) has a simpler expression,
\begin{eqnarray}
0 & = & \frac{U^{2}}{g^{rr}}\left(-\bar{R}_{ij}+KK_{ij}-K_{ic}K_{j}^{c}\right)+\left(1-\frac{U}{g^{rr}}\right)^{2}\left(\bar{R}_{titj}-K_{tt}K_{ij}+K_{tj}K_{ti}\right)\nonumber \\
 &  & +\frac{U}{\sqrt{g^{rr}}}\left(1-\frac{U}{g^{rr}}\right)\left(2D_{(j}K_{i)t}-2D_{t}K_{ij}\right)+O(\lambda^6).\label{eq:PetrovKij}
\end{eqnarray}
The induced metric of $\Sigma_c$ in new coordinate is
\begin{align}
ds_{p+1}^{2} & =\frac{g_{tt}}{4\hvar^2\lambda^{4}}d\tau^{2}+2\frac{g_{ti}}{2\hvar\lambda^{2}}d\text{\ensuremath{\tau}}dx^{i}+g_{ij}dx^{i}dx^{j}.\label{eq:inducemetric}
\end{align}
In coordinate $(\tau,r,x^{i})$, the Petrov-like condition (\ref{eq:PetrovKij})
becomes
\begin{eqnarray}
0 & = & \frac{U^{2}}{g^{rr}}\left(KK_{j}^{i}-K_{c}^{i}K_{j}^{c}-\bar{R}_{kj}h^{ki}\right)\nonumber\\
 &  & +4\hvar^2\lambda^{4}\left(1-\frac{U}{g^{rr}}\right)^{2}\left(\bar{R}_{\text{\ensuremath{\tau}}k\tau j}h^{ki}-K_{\tau\tau}K^i_{j}+K_{\tau j}K_{\tau}^i\right)\nonumber \\
 &  & +\frac{U}{\sqrt{g^{rr}}}\left(1-\frac{U}{g^{rr}}\right)2\hvar\lambda^{2}h^{ki}\left(2D_{(j}K_{k)\tau}-2D_{\tau}K_{kj}\right)+O(\lambda^6).\label{eq:Petrovtaui}
\end{eqnarray}
where $D_i$ is the induced connection on $\Sigma_c$.

According to the fluid/gravity correspondence, the stress tensor of
a gravity system corresponds to the energy-momentum tensor of a
fluid. {It is well known that for gravity such a stress tensor can
be described by the Brown-York tensor $t_{ab}$ }
\begin{eqnarray}
t_{ab}= Kh_{ab}-K_{ab}.
\end{eqnarray}
{We remark that} in the standard AdS/CFT {framework}, there should
be a counterterm in addition to the bare Brown-York tensor. However,
in our approach the counterterm will not affect the final result. A
short {argument} is given in the section of Summary and Discussion.
Here we just ignore it. Plugging into Eq.(\ref{eq:Petrovtaui}), we
get the Petrov-like conditions in terms of the stress tensor,
\begin{eqnarray}
0 & = & -\frac{U^{2}}{g^{rr}}\bar{R}_{kj}h^{ki}+4\hvar^2\lambda^{4}\left(1-\frac{U}{g^{rr}}\right)^{2}\bar{R}_{\text{\ensuremath{\tau}}i\tau j}h^{ki}\nonumber\\
&&+\frac{U^{2}}{g^{rr}}\left(-t_{\tau}^{i}t_{j}^{\tau}+\frac{t}{p}t_{j}^{i}-t_{k}^{i}t_{j}^{k}\right)\nonumber\\
 &  & -4\hvar^2\lambda^{4}\left(1-\frac{U}{g^{rr}}\right)^{2}\left[h_{\tau\tau}\left(\frac{t^{2}}{p^{2}}\delta_{j}^{i}-t_{\tau}^{\tau}\frac{t}{p}\delta_{j}^{i}-\frac{t}{p}t_{j}^{i}+t_{\tau}^{\tau}t_{j}^{i}-t_{j}^{\tau}t_{\tau}^{i}\right)\right.\nonumber \\
 &  & \ \ \left. +h_{k\tau}\left(\frac{t}{p}t_{\tau}^{i}\delta_{j}^{k}-t_{j}^{k}t_{\tau}^{i}+\frac{t}{p}t_{\tau}^{k}\delta_{j}^{i}-t_{\tau}^{k}t_{j}^{i}\right)
 \right]\nonumber \\
 &  & +\frac{U}{\sqrt{g^{rr}}}\left(1-\frac{U}{g^{rr}}\right)2\hvar\lambda^{2}h^{ki}\left[-2h_{\tau\tau}D_{(j}t_{k)}^{\tau}+2h_{\tau m}\left(D_{(j}\frac{t}{p}\delta_{k)}^{m}-D_{(j}t_{k)}^{m}\right)\right.\nonumber \\
 &  & \ \ \left.+2h_{k\tau}D_{\tau}t_{j}^{\tau}-2h_{km}D_{\tau}\left(\frac{t}{p}\delta_{j}^{m}-t_{j}^{m}\right)\right]+O(\lambda^6).\label{eq:PetrovTab}
\end{eqnarray}
Now we {have obtained} the expression of Petrov-like boundary
condition in terms of intrinsic geometry and extrinsic curvature of
$\Sigma_c$. Such boundary condition will provide some restriction
{on the relation} between the intrinsic geometry and extrinsic
curvature of $\Sigma_c$. Based on the requirement of fluid/gravity
correspondence, such boundary condition should also hold when
gravitational perturbation is introduced. In order to get the
concrete form of such restrictions, the $\lambda$-extension of right
hand side of Eq.(\ref{eq:PetrovTab}) is needed. In next subsection,
we will analysis the $\lambda$-extension of Brown-York tensor, then
get the $\lambda$-extension of terms in Eq.(\ref{eq:PetrovTab})
which is associated with Brown-York tensor. In subsection 3.2, we
will analysis the $\lambda$-extension of terms which contains
intrinsic curvature. Combining these results, one can get the
restrictions on perturbed Brown-York tensor which is caused by the
Petrov-like boundary condition. In section 4, we will see such
restriction will help us to get the dual Navier-Stocks equations.

\subsection{The order of the Brown-York tensor and constraints for perturbation}
First of all, the near horizon behavior of the extrinsic curvature
is needed.
\begin{equation}
K_{ab}=\frac{1}{2}\mathcal{L}_{N}h_{ab}=\frac{1}{\sqrt{g^{rr}}}\Gamma^r_{ab},\quad a,b=t,i
\end{equation}
where $N^{a}$ and $h_{ab}$ are the normal vector and the induced
metric of $\Sigma_{c}$ respectively. The components of $N^{a}$ are
\begin{equation}
N^{r}=\sqrt{g^{rr}}\ ,\quad N^{t}=\frac{1}{\sqrt{g^{rr}}}\ ,\quad N^{i}=\frac{g^{ri}}{\sqrt{g^{rr}}}\ .
\end{equation}
With the help of Eq.(\ref{eq:-gij}) and (\ref{gij}), the near horizon behaviour of $N^a$ is clear.

\begin{eqnarray}
\sqrt{g^{rr}} & = & \sqrt{2\hvar r}\left(1+\frac{\hR_{nlnl}}{4\hvar}r+O(r^2)\right)= \left({2\hvar}\lambda+4\hvar^2b\lambda^{3}\right)+O\left(\lambda^{5}\right),\nonumber \\
g^{ri} & = & \hR_{nInl}e_{I}^{i}r^{2}+O\left(r^{3}\right)= 4\hvar^2c^{i}\lambda^{4}+O\left(\lambda^{6}\right),\nonumber
\end{eqnarray}
where $b=\hR_{nlnl}/4\hvar$, $c^{i}=\hR_{nInl}e_{I}^{i}$.

The induced metric is easy to work out,
\begin{eqnarray}
&&h_{\text{\ensuremath{\tau\tau}}}=\frac{g_{tt}}{4\hvar^2\lambda^{4}}=-\frac{1}{\lambda^{2}}-\hR_{lnln}+O\left(\lambda^{2}\right) ,\qquad  h^{\tau\tau}=-\frac{\lambda^{4}}{g^{rr}}=-{\lambda^{2}}+O\left(\lambda^{4}\right);\nonumber \\
&&h_{\tau i}=\frac{g_{ti}}{2\hvar\lambda^{2}}\sim O\left(\lambda^{2}\right) ,\qquad  h^{\tau i}=-2\hvar\lambda^{2}\frac{g^{ri}}{g^{rr}}\sim O\left(\lambda^{4}\right);\nonumber\\
&&h_{ij}=g_{ij}\sim O\left(\lambda^{0}\right) ,\qquad
h^{ij}=g^{ij}-\frac{g^{ri}g^{rj}}{g^{rr}}=g^{ij}+O(\lambda^6).\label{h}
\end{eqnarray}
For {late} use, $\partial_{t}h_{ij}$ need to be calculated {out
explicitly}. From Eq.(\ref{1o}) and (\ref{eq:-gij}),
\begin{eqnarray}
g^{ij} & = & \hg^{ij}+2\he_{I}^{i}\widehat{(\partial_{r}e_{I}^{j})}r+O\left(r^{2}\right)\nonumber \\
 & = & \hg^{ij}-2\he_{I}^{i}{\hat\theta}'_{JI}\he_{J}^{j}r+O\left(r^{2}\right),\\
g_{ij} & = & \hg_{ij}+\hg_{ik}\he_{I}^{k}{\hat\theta}'_{JI}\he_{J}^{k}\hg_{kj}r+O\left(r^{2}\right).\nonumber
\end{eqnarray}
Weakly isolated horizon condition implies $\partial_{t}\hg_{ij}=0$. Thus
\begin{eqnarray}
\partial_{t}g_{ij}  =  [2\hg_{ik}(\partial_{t}\he_{I}^{k}){\hat\theta}'_{JI}\he_{J}^{m}\hg_{mj}+\hg_{ik}\he_{I}^{k}(\partial_{t}\hth'_{JI})\he_{J}^{m}\hg_{mj}]r+O\left(r^{2}\right).\label{eq:gijdt}
\end{eqnarray}
We need the values of $\partial_{t}e_{I}^{k}$ and $\partial_{t}\theta'_{JI}$
on horizon. With the help of facts $X^{j}\heq U\heq W_I\heq 0$, the commutator between $l$ and $E_I$ gives
\begin{eqnarray}
\left[l,E_{I}\right] & \hat{=} & (\partial_{t}e_{I}^{i})\partial_{i}.
\end{eqnarray}
On the other hand,
\begin{eqnarray}
\left[l,E_{I}\right] & = & \nabla_{l}E_{I}-\nabla_Il\\
 & = & -\pi_{I}l+\varepsilon_{JI}E_{J}-\alpha_{I}l-\theta_{JI}E_{J}.\nonumber
\end{eqnarray}
Here we use abbreviations $\varepsilon_{JI}=\left\langle E_{J},\nabla_{l}E_{I}\right\rangle ,\theta_{JI}=\left\langle E_{J},\nabla_Il\right\rangle $.
The fact that on horizon $l$ is geodesic leads to $\left\langle l,\nabla_{l}E_{I}\right\rangle =-\left\langle E_{I},\nabla_{l}l\right\rangle\heq 0$.
The Bondi gauge of the null tetrad and the fact that $l$ is normal direction of horizon implies $\varepsilon_{IJ}\heq 0$.
For WIH condition $tr\theta\heq 0$, it implies that $\theta{}_{IJ}\heq0$
by Raychaudhuri equation. Thus we get $\left[l,E_{I}\right]\heq0$ for non-rotating black hole.
This implies that $\partial_{t}e_{I}^{i}\heq0$.

As for $\partial_{t}\theta'_{JI}$, consider the derivation
\begin{eqnarray}
\nabla_{l}\theta'_{JI} & = & \nabla_{l}\left\langle E_{J},\nabla_In\right\rangle \nonumber\\
 & \hat{=} & R_{lIJn}-\varepsilon\theta'_{JI},
\end{eqnarray}
where the fact $\nabla_{I}\pi_{J}\heq0$, $\pi_{J}\heq\alpha_{I}\heq\theta{}_{JK}\heq\varepsilon_{KI}\heq0$
and $U\heq X^{i}\heq0$ are used. On the other hand, because of Eq.(\ref{tetrad0})
\begin{eqnarray}
\nabla_{l}\theta'_{JI} & \hat{=} & \partial_{t}\theta'_{JI}.
\end{eqnarray}
Thus we have
\begin{eqnarray}
\partial_{t}\theta'_{JI} &\hat{=}& R_{lIJn}-\varepsilon\theta'_{JI}.
\end{eqnarray}
Plugging into Eq.(\ref{eq:gijdt}), we may deduce that
\begin{eqnarray}
\partial_{t}g_{ij}=\hg_{ik}\he_{I}^{k}(\partial_{t}\hth'_{JI})\he_{J}^{m}\hg_{mj}2\hvar\lambda^2+O(\lambda^4)  \sim O(\lambda^{2}).
\end{eqnarray}
We also need the value of $\partial_rg_{ij}$. Because $g_{ik}g^{kj}=\delta^j_i$ and $g^{ij}=\sum e^i_Ie^j_I$, it is easy to see
\begin{eqnarray}
\partial_rg_{ij}&=&-g_{mi}(\partial_re^m_I)e^k_Ig_{kj}-g_{mj}(\partial_re^m_I)e^k_Ig_{ki}\nonumber\\
&=&2g_{mi}\theta'_{IJ}e^m_Je^k_Ig_{kj}\sim O(\lambda^0).
\end{eqnarray}
In the last step, Eq.(\ref{1o}) {is} used.

Based on these preparation, the near horizon extension of {the}
extrinsic curvature are
\begin{eqnarray}
K_{j}^{i} & = & h^{\mu i}K_{\mu j}=h^{ki}K_{kj}+h^{ti}K_{tj}\nonumber \\
 & = & \frac{1}{2}g^{ki}\left(\sqrt{g^{rr}}\partial_{r}g_{kj}+\frac{1}{\sqrt{g^{rr}}}\partial_{t}g_{kj}\right)+O\left(\lambda^{3}\right)\nonumber \\
 & = & \sqrt{2\hvar}\xi_{j}^{i}\lambda+O\left(\lambda^{3}\right)\sim O\left(\lambda\right).\label{eq:Kij}\\
K_{i}^{\tau} & = & 2\hvar\lambda^{2}K_{i}^{t}= \hvar\lambda^{2}(h^{tt}K_{ti}+h^{kt}K_{ki})\sim O(\lambda^4)\\
K^{\tau}_{\tau}&=&K_{t}^{t}=h^{tt}K_{tt}+h^{it}K_{it}\nonumber \\
 & = & -\frac{1}{2g^{rr}}\left(\sqrt{g^{rr}}\partial_{r}g_{tt}+\frac{1}{\sqrt{g^{rr}}}\partial_{t}g_{tt}+2\partial_{t}\sqrt{g^{rr}}+2g_{tt}\partial_{t}\frac{1}{\sqrt{g^{rr}}}\right)+O\left(\lambda^{3}\right)\nonumber \\
 & = & \frac{1}{2\lambda}+\beta\lambda+O\left(\lambda^{3}\right)\sim O\left(\lambda^{-1}\right).\\
K & = & K_{t}^{t}+K_{i}^{i}\nonumber \\
 & = & \frac{1}{2\lambda}+\sqrt{2\hvar}(\beta+\xi)\lambda+O\left(\lambda^{3}\right)\sim O(\frac{1}{\lambda}).\label{eq:Ksi}
\end{eqnarray}
Here $\xi_{j}^{i}= \hg^{ki}[\sqrt{2\hvar}\
\hat{\theta}'_{IJ}+(2\hvar)^{-1/2}\partial_t\hat{\theta}'_{IJ}]\he^I_k\he^J_j$
is the coefficient of the leading order of $K_{j}^{i}$, {while }
$\beta$ is the cofficient of the order $O\left(\lambda\right)$ in
$K_{t}^{t}$ and $\xi=\xi_{i}^{i}$.

{We remark that} Eq.(\ref{eq:Ksi}) shows the mean curvature of
$\Sigma_c$ is $K\sim O(\frac{1}{\lambda})$, which is similar to the
large mean curvature expansion {obtained} by Lysov and Strominger in
Rindler case\cite{LS-PE2NS}. This means up to the order we will use
LS's large mean curvature limit is equivalent to our near horizon
limit.

Now let us introduce the gravitational perturbations. Based on main
picture of AdS/CFT correspondence, the intrinsic geometry of
boundary is fixed and the extrinsic curvature of boundary is
perturbed. Obviously, perturbing the extrinsic curvature is
equivalent to perturbing the Brown-York tensor. As done by
\cite{LS-PE2NS},  the concrete form of the perturbation is
\begin{eqnarray}
t^a_b={}^Bt^a_b+\sum_{k=1} t^{a(k)}_b\lambda^k,
\end{eqnarray}
where ${}^Bt^a_b$ is the Brown-York tensor of back ground spacetime.

As {stressed} before, the Petrov-like boundary condition plays an
extremely important rule in fluid/gravity correspondence. In terms
of initial-boundary value problem of Einstein system\cite{IBVP},
such boundary condition means there is no out-going gravitational
radiation through boundary. Such boundary condition also should hold
for perturbed {spacetime} and give some restrictions for the gravity
perturbation. Let {us} consider the Petrov-like boundary condition
for perturbed spacetime. Remember the near horizon {behavior} of
$K^a_b$, i.e. Eq.(\ref{eq:Kij})-(\ref{eq:Ksi}), the perturbed
Brown-York tensor is
\begin{eqnarray}
t_{i}^{\tau} & = & -K_{i}^{\tau}+\sum_{k=1} t^{a(k)}_b\lambda^k\nonumber\\
 & = & \lambda t_{i}^{\tau(1)}+O\left(\lambda^{2}\right), \\
t_{\tau}^{\tau} & = & K-K_{\tau}^{\tau}+\sum_{k=1} t^{a(k)}_b\lambda^k=K_{i}^{i}+\sum_{k=1} t^{a(k)}_b\lambda^k\nonumber \\
 & = & \left(\sqrt{2\hvar}\xi+t_{\tau}^{\tau(1)}\right)\lambda+O\left(\lambda^{2}\right),\\
t_{j}^{i} & = & K\delta_{j}^{i}-K{}_{j}^{i}+\sum_{k=1} t^{a(k)}_b\lambda^k\nonumber \\
 & = & \frac{1}{2\lambda}\delta_{j}^{i}+\left\{\sqrt{2\hvar}\ [(\beta+\xi)\delta_{j}^{i}-\xi_{j}^{i}]+t_{j}^{i(1)}\right\}\lambda+O\left(\lambda^{2}\right), \\
t^i_{\tau}&=&h^i_{\mu}h^{\nu}_{\tau}t^{\mu}_{\nu}=h^{\tau i}h_{\tau\tau}t^{\tau}_{\tau}+h^{\tau i}h_{j\tau}t^{j}_{\tau}+h^{j i}h_{\tau\tau}t^{\tau}_{j}
+h^{j i}h_{m\tau}t^{m}_{j}\nonumber\\
&=&h^{j i}h_{\tau\tau}t^{\tau}_{j}+O(\lambda)\nonumber\\
&=&-\frac{1}{\lambda}h^{ij}t^{\tau(1)}_j+O(1),  \\
t & = & pK+\sum_{k=1} t^{a(k)}_b\lambda^k\nonumber\\
 & = &\frac{p}{2\lambda}+\left(p\sqrt{2\hvar}\ (\beta+\xi)+t^{(1)}\right)\lambda+O\left(\lambda^{2}\right).
\end{eqnarray}
{Substituting} these results into the Petrov-like condition
(\ref{eq:PetrovTab}), we want to show that $t_{j}^{i(1)}$ could be
expressed in terms of $t_{i}^{\tau(1)}$. The coefficients of
Eq.(\ref{eq:PetrovTab}) are
\begin{eqnarray}
\frac{U^{2}}{g^{rr}} & = & {\hvar}^{2}\lambda^{2}+O\left(\lambda^{4}\right),\nonumber \\
-\left(1-\frac{U}{g^{rr}}\right)^{2}4\hvar^2\lambda^{4} & = & -\hvar^2\lambda^{4}+O\left(\lambda^{6}\right),\label{eq:OrderofCoff}\\
\frac{U}{\sqrt{g^{rr}}}\left(1-\frac{U}{g^{rr}}\right)2\hvar\lambda^{2} & = &\hvar^2\lambda^{3}+O\left(\lambda^{5}\right).\nonumber
\end{eqnarray}
The terms in parentheses of Eq.(\ref{eq:PetrovTab}) which are related to the Brown-York tensor
could be calculated straightly,
\begin{eqnarray}
&&-t_{\tau}^{i}t_{j}^{\tau}+\frac{t}{p}t_{j}^{i}-t_{k}^{i}t_{j}^{k} \nonumber\\
&=& h^{ik}t_{k}^{\tau(1)}t_{j}^{\tau(1)}+\frac{1}{2}\left(-t_{j}^{i(1)}+\frac{t^{(1)}}{p}\delta_{j}^{i}\right)+\sqrt{\frac{\hvar}{2}}\ \xi^i_j+O\left(\lambda\right),\nonumber\\
&&\frac{t^{2}}{p^{2}}\delta_{j}^{i}-t_{\tau}^{\tau}\frac{t}{p}\delta_{j}^{i}-\frac{t}{p}t_{j}^{i}+t_{\tau}^{\tau}t_{j}^{i}-t_{j}^{\tau}t_{\tau}^{i}\nonumber\\
&=& \frac{1}{2}\left(\frac{t^{(1)}}{p}\delta_{j}^{i}-t_{j}^{i(1)}\right)+\sqrt{\frac{\hvar}{2}}\ \xi_{j}^{i}+h^{ik}t_{k}^{\tau(1)}t_{j}^{\tau(1)}+O\left(\lambda\right),\nonumber \\
&&\lambda^{2}h_{k\tau}\left(\frac{t}{p}t_{\tau}^{i}\delta_{j}^{k}-t_{j}^{k}t_{\tau}^{i}-\frac{t}{p}t_{\tau}^{k}\delta_{j}^{i}+t_{\tau}^{k}t_{j}^{i}\right) \sim O\left(\lambda^{2}\right).
\end{eqnarray}

Then we consider the derivations $D_{a}t_{c}^{b}$. In order to do that, we need
the connection coefficients ${\bar\Gamma}_{bc}^{a}$ on $\Sigma_{c}$.
Since the induced metric on $\Sigma_c$ is fixed, by definition, ${\bar\Gamma}_{bc}^{a}=\frac{1}{2}h^{ad}\left(h_{bd,c}+h_{cd,b}-h_{bc,d}\right)$.
Straightforward calculation shows
\begin{eqnarray}
{\bar\Gamma}_{\tau\tau}^{\tau} & = & h^{\tau\tau}h_{\tau\tau,\tau}+O\left(\lambda^{2}\right)=\frac{1}{2\hvar}\partial_t\hR_{nlnl}+O(\lambda^2)\sim O\left(\lambda^{0}\right),\nonumber \\
{\bar\Gamma}_{j\tau}^{\tau} & = & \frac{1}{2}h^{\tau\tau}h_{\tau\tau,j}+O\left(\lambda^{4}\right)\sim O\left(\lambda^{2}\right),\nonumber \\
{\bar\Gamma}_{ji}^{\tau} & = & -\frac{1}{2}h^{\tau\tau}h_{ji,\tau}+O\left(\lambda^{4}\right)\sim O\left(\lambda^{2}\right),\nonumber\\
{\bar\Gamma}_{\tau\tau}^{i} & = & \frac{1}{2}h^{ik}\left(2h_{\tau k,\tau}-h_{\tau\tau,k}\right)+O\left(\lambda^{2}\right)\sim O\left(\lambda^{0}\right),\nonumber \\
{\bar\Gamma}_{j\tau}^{m} & = & \frac{1}{2}h^{mk}h_{jk,\tau}+O\left(\lambda^{2}\right)=\frac{1}{2}\hg_{ji}\he^i_I(\partial_t\hth'_{IJ})\he^m_J+O(\lambda^2)\sim O\left(\lambda^{0}\right),\nonumber \\
{\bar\Gamma}_{jk}^{i}&=&\frac{1}{2}h^{im}\left(h_{jm,k}+h_{km,j}-h_{jk,m}\right)+O\left(\lambda^{4}\right)\nonumber\\
&=&\tGa^i_{jk}+O(\lambda^2)\sim
O\left(\lambda^{0}\right),\label{eq:Gamma}
\end{eqnarray}
where
$\tGa^i_{jk}=\frac{1}{2}\hg^{im}(\hg_{jm,k}+\hg_{km,j}-\hg_{jk,m})$
is the Christoffol symbol on the section $S_t$ of horizon.

Based on these preparations, we get
\begin{eqnarray}
&&-\frac{U}{\sqrt{g^{rr}}}\left(1-\frac{U}{g^{rr}}\right)2\varepsilon\lambda^{2}h^{ki}2h_{\tau\tau}D_{(j}t_{k)}^{\tau} = -2\hvar^2\lambda^{2}h^{ki}D_{(j}t_{k)}^{\tau(1)}+O\left(\lambda^{4}\right),\nonumber \\
&&\frac{U}{\sqrt{g^{rr}}}\left(1-\frac{U}{g^{rr}}\right)2\varepsilon\lambda^{2}h^{ki}2h_{\tau m}\left(D_{(j}\frac{t}{p}\delta_{k)}^{m}-D_{(j}t_{k)}^{m}\right)  \sim  O\left(\lambda^{4}\right),\nonumber\\
&&\frac{U}{\sqrt{g^{rr}}}\left(1-\frac{U}{g^{rr}}\right)2\varepsilon\lambda^{2}h^{ki}2h_{k\tau}D_{\tau}t_{k}^{\tau}  \sim  O\left(\lambda^{4}\right),\nonumber \\
&&-\frac{U}{\sqrt{g^{rr}}}\left(1-\frac{U}{g^{rr}}\right)2\varepsilon\lambda^{2}h^{ki}2h_{km}D_{\tau}\left(\frac{t}{p}\delta_{j}^{m}-t_{j}^{m}\right)\nonumber\\
&&\qquad = O\left(\lambda^{2}\right)\sim O\left(\lambda^{0}\right).
\end{eqnarray}

Now most parts of Eq.(\ref{eq:PetrovTab}) have been written out.
Only two terms which are related to the intrinsic curvature are remained.

\subsection{Order of the intrinsic curvature}

In previous subsection, we calculate all terms in Eq.(\ref{eq:PetrovTab}) which contains the perturbed Brown-York tensor. There still two terms left, i.e. the two terms in the first line of Eq.(\ref{eq:PetrovTab}). Each of them contains Riemannian curvature of the boundary $\Sigma_c$. Now we consier these two terms

\begin{equation}
-\frac{U^{2}}{g^{rr}}\bar{R}_{kj}h^{ki}+4\hvar^2\lambda^{4}\left(1-\frac{U}{g^{rr}}\right)^{2}\bar{R}_{\text{\ensuremath{\tau}}i\tau j}h^{ki}.
\end{equation}

By definition,
\begin{eqnarray}
\bar{R}_{\tau k\tau j} & = & h_{\mu j}\bar{R}_{\tau k\tau}^{\ \ \ \mu}=h_{\tau j}\bar{R}_{\tau k\tau}^{\ \ \ \ \tau}+h_{ij}\bar{R}_{\tau k\tau}^{\ \ \ \ i},\\
\bar{R}_{\tau k\tau}^{\ \ \ \ \tau} & = & \bar{\Gamma}_{\tau\tau,k}^{\tau}-\bar{\Gamma}_{\tau k,\tau}^{\tau}+\bar{\Gamma}_{\tau\tau}^{\mu}\bar{\Gamma}_{\mu k}^{\tau}-\bar{\Gamma}_{\tau k}^{\mu}\bar{\Gamma}_{\tau\mu}^{\tau}\nonumber \\
 & \sim & O\left(\lambda^{0}\right),\\
\bar{R}_{\tau k\tau}^{\ \ \ \ i} & = & \bar{\Gamma}_{\tau\tau,k}^{i}-\bar{\Gamma}_{\tau k,\tau}^{i}+\bar{\Gamma}_{\tau\tau}^{\mu}\bar{\Gamma}_{\mu k}^{i}-\bar{\Gamma}_{\tau k}^{\mu}\bar{\Gamma}_{\tau\mu}^{i}\nonumber \\
 & = & -\bar{\Gamma}_{\tau k,\tau}^{i}+O\left(\lambda^{0}\right)\nonumber\\
 & \sim & O\left(\lambda^{-2}\right),
\end{eqnarray}
with Eq.(\ref{eq:Gamma}),
\begin{eqnarray}
4\hvar^2\lambda^{4}\left(1-\frac{U}{g^{rr}}\right)^{2}\bar{R}_{\text{\ensuremath{\tau}}i\tau
j}h^{ki}\sim O(\lambda^2).\label{1}
\end{eqnarray}
For Ricci tensor of $\Sigma_{c}$,
\begin{eqnarray}
\bar{R}_{kj} & = & \bar{R}_{k\mu j}^{\ \ \ \mu}=\bar{R}_{k\tau j}^{\ \ \ \tau}+\bar{R}_{kij}^{\ \ \ i},\\
\bar{R}_{k\tau j}^{\ \ \ \tau} & = & \bar{\Gamma}_{kj,\tau}^{\tau}-\bar{\Gamma}_{\tau j,k}^{\tau}+\bar{\Gamma}_{kj}^{\mu}\bar{\Gamma}_{\mu\tau}^{\tau}-\bar{\Gamma}_{\tau j}^{\mu}\bar{\Gamma}_{k\mu}^{\tau}\nonumber \\
 & = & \bGa_{kj,\tau}^{\tau}+O\left(\lambda^{2}\right)\nonumber \\
 &=&\frac{1}{2\hvar}\ (\partial^2_t\hth'_{IJ})\he^J_j\he^I_k+O(\lambda^2)\nonumber\\
 &\sim&  O\left(\lambda^{0}\right),\\
\bar{R}_{kij}^{\ \ \ i} & = & \bar{\Gamma}_{kj,i}^{i}-\bar{\Gamma}_{ij,k}^{i}+\bar{\Gamma}_{kj}^{\mu}\bar{\Gamma}_{\mu i}^{i}-\bar{\Gamma}_{ij}^{\mu}\bar{\Gamma}_{k\mu}^{i}\nonumber \\
 & = & \bar{\Gamma}_{kj,i}^{i}-\bar{\Gamma}_{ij,k}^{i}+\bar{\Gamma}_{kj}^{m}\bar{\Gamma}_{mi}^{i}-\bar{\Gamma}_{ij}^{m}\bar{\Gamma}_{km}^{i}+O\left(\lambda^{2}\right)\nonumber \\
 & = & \tilde{R}_{kj}+O\left(\lambda^{2}\right)\nonumber\\
 &\sim&  O\left(\lambda^{0}\right),
\end{eqnarray}
where $\tilde{R}_{j}^{i}$ is the Riemann curvature of $S_t$. It is
obvious that $\tR_{kj}\sim O(\lambda^0)$. Then it is easy to see
that
\begin{eqnarray}
&&-\frac{U^{2}}{g^{rr}}\bar{R}_{kj}h^{ki}\nonumber\\
&=&-\left[\hvar^2\lambda^2+{\hR_{nlnl}}\hvar^2\lambda^4+O(\lambda^6)\right]\bar{R}_{kj}h^{ki}\nonumber\\
&=&-\left[\hvar^2\lambda^2+{\hR_{nlnl}}\hvar^2\lambda^4+O(\lambda^6)\right]
\left[\frac{1}{2\hvar}\ (\partial^2_t\hth'_{IJ})\he^J_j\he^I_k+\tilde{R}_{kj}+O(\lambda^{2})\right]\nonumber\\
&=&-\frac{\hvar}{2}\ (\partial^2_t\hth'_{IJ})\he^J_j\he^I_k\lambda^2-{\hvar}^2\lambda^2\tilde{R}_{kj}+O(\lambda^4)\label{2}
\end{eqnarray}

Besides, straight calculation also shows that
\begin{eqnarray}
\left(-\lambda^{2}\bar{R}_{kj}+\lambda^{4}\bar{R}_{\text{\ensuremath{\tau}}i\tau j}\right)h^{ki}+\lambda^{3}D_{\tau}\left(\frac{t}{p}\delta_{j}^{i}-t_{j}^{i}\right)  =  -\lambda^{2}\tilde{R}_{j}^{i}+O\left(\lambda^{2}\right)
\end{eqnarray}

All the terms in Petrov-like condition has been worked out now. Combining
these results together, the 0-order of
Eq.(\ref{eq:PetrovTab}) holds automatically because the background fulfills the Petrov-like condition. The 2-order term gives
\begin{equation}
t_{j}^{i(1)}=2h^{ik}t_{k}^{\tau(1)}t_{j}^{\tau(1)}-2h^{ik}\tna_{\text{(}j}t_{k)}^{\tau(1)}+\frac{t^{(1)}}{p}\delta_{j}^{i}+\xi_{j}^{i}\sqrt{2\hvar}-\tilde{R}_{j}^{i}.\label{eq:tij}
\end{equation}
(Note : in above equation, the meaning of $\tna_it^{\tau(1)}_j$ is
that we see $t^{\tau(1)}_j$ as an 1-form on $S_t$ and take covariant
derivative on that vector inside $S_t$.) Here we see that the
Petrov-like boundary condition gives restriction for the perturbed
Brown-York tensor, i.e. $t^{i(1)}_j$ are functions of $t^{i(1)}_j$
and other geometric quantities on horizon. This restriction is
similar to what has been gotten by previous works for concrete
space-time back
ground\cite{LS-PE2NS,HLPTW_PE2NS-curvedspatial,HLPTW_PE2NS-cosmol,ZLNTW_MG}.

\subsection{Gaussian Equation}
Besides the Petrov-like boundary condition, there is another
constrain equation on $\Sigma_c$, i.e. the Gaussian equation, which
also can help us to restrict the freedom of perturbation of
Brown-York tensor. The Gaussian equation for gravity takes the form

\[
\bar{R}+K^{\mu\nu}K_{\mu\nu}-K^{2}=0.
\]
In terms of the Brown-York tensor $t_{ab}$, it becomes
\[
\bar{R}+\left(t_{\tau}^{\tau}\right)^{2}-\frac{2h^{ij}}{\lambda^{2}}t_{i}^{\tau}t_{j}^{\tau}-\frac{t^{2}}{p}+t_{j}^{i}t_{i}^{j}=0.
\]
Expanding the scalar curvature $\bR$ in terms of parameter $\lambda$
near the horzion, we have
\begin{eqnarray*}
\bar{R} & = & \bar{R}_{\mu\nu}h^{\mu\nu}=\bar{R}_{\tau\tau}h^{\tau\tau}+2\bar{R}_{\tau i}h^{\tau i}+\bar{R}_{ij}h^{ij}\\
 & = & -\frac{1}{2g_{tt}}h^{im}\partial_{t}\partial_{t}h_{im}+\tilde{R}+O(\lambda^{2})\\
 & \sim & O(\lambda^{0})
\end{eqnarray*}
Then 0-order component of Gaussian equation gives
\begin{eqnarray}
t_{\tau}^{\tau(1)}=\bar{R}-2h^{ij}t_{i}^{\tau(1)}t_{j}^{\tau(1)}-\sqrt{\frac{\hvar}{2}}\ \xi,
\end{eqnarray}
where $\xi$ is a parameter introduced in Eq.(\ref{eq:Ksi}). Similar
to previous works, the Gaussian equation help us to fix the
perturbation $t^{\tau(1)}_{\tau}$ in terms of $t^{\tau(1)}_j$ and
horizon geometry. This equation could be viewed as the equation of
state for dual fluid.

\section{Navier-Stokes equation}
In previous section, we {have considered} the large mean curvature
extension of the Petrov-like boundary condition for general
non-rotating black {holes}. Like the discussion by Lysov and
Strominger\cite{LS-PE2NS}, the Petrov-like condition reduces the
degree of freedom of the gravity to those of a fluid. The remaining
$p+2$ variables may be interpreted as the hydrodynamic variables of
a fluid living on the boundary. The Gaussian equation for gravity
then can be viewed as the equation of state. In this section, we
consider the Codazzi equation on $\Sigma_c$ and show it will {lead}
to the Navier-Stokes equation {at  the  large mean curvature limit}.

The Codazzi equation in terms of Brown-York tensor is
\begin{eqnarray}
D_{a}t_{b}^{a}=0
\end{eqnarray}
Because the background spacetime satisfies the Codazzi equation
automatically, we only need to consider the equation for
perturbation.
\begin{eqnarray}
0&=&D_a\sum_{k=1} t^{a\tau(k)}\lambda^k \nonumber\\
&=&\partial_{\tau}\sum_{k=1}
t^{\tau\tau(k)}\lambda^k+\partial_i\sum_{k=1}
t^{i\tau(k)}\lambda^k+\bGa_{\tau\tau}^{\tau}\sum_{k=1}
t^{\tau\tau(k)}\lambda^k+\bGa_{i\tau}^{i}\sum_{k=1}
t^{\tau\tau(k)}\lambda^k\nonumber\\
&&+\bGa_{\tau i}^{\tau}\sum_{k=1}
t^{i\tau(k)}\lambda^k+\bGa_{ij}^{i}\sum_{k=1}
t^{j\tau(k)}\lambda^k\nonumber\\
&&+\bGa_{\tau\tau}^{\tau}\sum_{k=1}
t^{\tau\tau(k)}\lambda^k+2\bGa_{\tau i}^{\tau}\sum_{k=1} t^{\tau
i(k)}\lambda^k+\bGa_{ij}^{\tau}\sum_{k=1} t^{ij(k)}\lambda^k
\end{eqnarray}
With {the} help of Eq.(\ref{h}), it is easy to see that
\begin{eqnarray}
t^{\tau\tau(1)} & = & h^{\tau\tau}t_{\tau}^{\tau(1)}+h^{i\tau}t_i^{\tau(1)}\sim O\left(\lambda^{2}\right),\nonumber \\
t^{i\tau(1)} & = & h^{i\tau}t_{\tau}^{\tau(1)}+h^{ij}t_j^{\tau(1)}=h^{ij}t_j^{\tau(1)}+O(\lambda^4)\sim O\left(\lambda^{0}\right),\\
t^{ij(1)} & = &
h^{i\tau}t_{\tau}^{j(1)}+h^{ik}t_k^{j(1)}=h^{ik}t_k^{j(1)}+O(\lambda^4)\sim
O\left(\lambda^{0}\right).\nonumber
\end{eqnarray}
Using Eq.(\ref{h}) and (\ref{eq:Gamma}),
\begin{eqnarray}
0&=&D_a\sum_{k=1} t^{a\tau(k)}\lambda^k \nonumber\\
&=&\left[\tna^jt_j^{\tau(1)}\right]\lambda+O(\lambda^2).\label{inc}
\end{eqnarray}
Based on the basic identification\cite{LS-PE2NS},
\begin{eqnarray}
2 t_{j}^{\tau(1)}\leftrightarrow v_{j},\qquad
2{t^{(1)}}/{p}\leftrightarrow P,\label{id}
\end{eqnarray}
under the large mean curvature limit, the leading order of
Eq.(\ref{inc}) implies $\tna^iv_i=0$, i.e. the incompressible
condition.

Similarly,
\begin{eqnarray}
0&=&D_a\sum_{k=1} t^{aj(k)}\lambda^k \nonumber\\
&=&\partial_{\tau}\sum_{k=1} t^{\tau
j(k)}\lambda^k+\partial_i\sum_{k=1}
t^{ij(k)}\lambda^k+\bGa_{\tau\tau}^{\tau}\sum_{k=1} t^{\tau
j(k)}\lambda^k+\bGa_{i\tau}^{i}\sum_{k=1}
t^{\tau j(k)}\lambda^k\nonumber\\
&&+\bGa_{\tau i}^{\tau}\sum_{k=1}
t^{ij(k)}\lambda^k+\bGa_{im}^{i}\sum_{k=1}
t^{mj(k)}\lambda^k\nonumber\\
&&+\bGa_{\tau\tau}^{j}\sum_{k=1}
t^{\tau\tau(k)}\lambda^k+2\bGa_{\tau i}^{j}\sum_{k=1} t^{\tau
i(k)}\lambda^k+\bGa_{mi}^{j}\sum_{k=1} t^{mi(k)}\lambda^k\nonumber\\
&=&[\partial_{\tau}(\hg^{ji}t^{\tau(1)}_i)+\partial_i(\hg^{ik}t^{j(1)}_k)+\bGa^{\tau}_{\tau\tau}t^{j\tau(1)}
+\bGa^{i}_{i\tau}t^{j\tau(1)}\nonumber\\
&&\quad+\tGa^{i}_{im}t^{mj(1)}+2\bGa^{j}_{i\tau}t^{i\tau(1)}+\tGa^{j}_{mi}t^{mi(1)}]\lambda+O(\lambda^2)\nonumber\\
&=&[\hg^{ji}\partial_{\tau}t^{\tau(1)}_i+\tna^i(t^{j(1)}_i)+\bGa^{\tau}_{\tau\tau}t^{j\tau(1)}
+\bGa^{i}_{i\tau}t^{j\tau(1)}
+2\bGa^{j}_{i\tau}t^{i\tau(1)}]\lambda+O(\lambda^2)\nonumber\\
\end{eqnarray}
Using Eq.(\ref{eq:tij}), above equation becomes
\begin{eqnarray}
0&=&[\hg^{ji}\partial_{\tau}t^{\tau(1)}_i+\tna^i(t^{j(1)}_i)+\bGa^{\tau}_{\tau\tau}t^{j\tau(1)}
+\bGa^{i}_{i\tau}t^{j\tau(1)}
+2\bGa^{j}_{i\tau}t^{i\tau(1)}]\lambda+O(\lambda^2)\nonumber\\
&=&\left[\hg^{ji}\partial_{\tau}t^{\tau(1)}_i+\tna^i\left(2\hg^{jk}t_{k}^{\tau(1)}t_{i}^{\tau(1)}-2
\hg^{jk}\tna_{(k}t_{i)}^{\tau(1)}
+\frac{t^{(1)}}{p}\delta_{i}^{j}+\xi_{i}^{j}\sqrt{2\hvar}-\tilde{R}_{i}^{j}\right)\right.\nonumber\\
&&\quad\left.+\bGa^{\tau}_{\tau\tau}t^{j\tau(1)}
+\bGa^{i}_{i\tau}t^{j\tau(1)}
+2\bGa^{j}_{i\tau}t^{i\tau(1)}\right]\lambda+O(\lambda^2)\nonumber\\
&=&\left[\hg^{ji}\partial_{\tau}t^{\tau(1)}_i +2
t^{\tau(1)}_i\tna^i(t^{j\tau(1)})+2
t^{j\tau(1)}\tna^i(
t^{\tau(1)}_i)\right.\nonumber\\
&&\quad-2\tna^i\hg^{jk}\tna_{(k}t^{\tau(1)}_{i)}+\tna^i\left(\frac{t^{(1)}}{p}\delta_{i}^{j}\right)+\tna^i(\xi_{i}^{j}\sqrt{2\hvar}-\tilde{R}_{i}^{j})\nonumber\\
&&\quad\left.+\bGa^{\tau}_{\tau\tau}t^{j\tau(1)}
+\bGa^{i}_{i\tau}t^{j\tau(1)}
+2\bGa^{j}_{i\tau}t^{i\tau(1)}\right]\lambda+O(\lambda^2).\nonumber
\end{eqnarray}
Then with Eq.(\ref{inc}) and (\ref{id}), we get the leading order of
above equation {as}
\begin{eqnarray}
0&=&\partial_{\tau}v^j
+v^i\tna_iv^j-\tilde{\triangle}v^j-\tR_i^jv^i
+\tna^jP+f^j\nonumber\\
&&+\bGa^{\tau}_{\tau\tau}v^{j} +\bGa^{i}_{i\tau}v^{j}
+2\bGa^{j}_{i\tau}v^{i}\label{eq:NS}
\end{eqnarray}
where $f^j=\tna^i(\xi_{i}^{j}\sqrt{2\hvar}-\tilde{R}_{i}^{j})$ only
depends on {the} horizon geometry, so it can be seen as an external
{force} caused by {the} curved space.

Remark : in Eq.(\ref{eq:NS}), we get a generalized Navier-Stocks
equation. The first line of eq.(\ref{eq:NS}) is in the form of
standard Navier-Stocks equation with external force. The problem is
the {extra} terms in {the} second line. These terms {are} related
with the Christoffel symbol on horizon. From eq.(\ref{eq:Gamma}), it
is easy to see that
\begin{eqnarray}
{\bar\Gamma}_{\tau\tau}^{\tau} &\heq& \frac{1}{2\hvar}\partial_t\hR_{nlnl},\nonumber \\
{\bar\Gamma}_{j\tau}^{i} &\heq& \frac{1}{2}\hg_{jk}\he^k_I(\partial_t\hth'_{IJ})\he^i_J,\nonumber \\
{\bar\Gamma}_{i\tau}^{i} &\heq& \frac{1}{2}\partial_t(\hth'_{IJ})\delta^{IJ}.
\end{eqnarray}
If the black hole is stationary, $\partial_t$ is {a} Killing vector.
Obviously, all these additional terms vanish, so we get the standard
Navier-Stocks equation. This means that fluid/gravity correspondence
exists for general non-rotating stationary black hole. Furthermore,
simple calculation shows that these additional terms will be zero if
$\partial_t$ satisfies $\cL_{\partial_t}g\sim O(r^2)$, so such
correspondence also holds for asymptotic stationary black holes
(Note: based on Eq.(\ref{eq:gijdt}), $\cL_{\partial_t}g_{ab}\sim
O(r)$ for general WIH, so our asymptotic stationary condition
$\cL_{\partial_t}g\sim O(r^2)$ is not very restricted.). This result
tells us that fluid/gravity correspondence can also exist in some
dynamical cases. Another remark is about the external force term. In
previous
works\cite{LS-PE2NS,BKLS11,HLPTW_PE2NS-curvedspatial,HLPTW_PE2NS-cosmol},
all cases satisfy $\xi^i_j=0$ and $\tna_i\tR^i_j=0$, so there are no
external force term and our result {agrees} with previous results.
For general horizon case, the {inhomogeneity}  of the section of
horizon will cause an external force term in the dual Navier-Stocks
equation.

\section{Summary and discussion}

In this paper, we have proved that Lysov and Strominger's
realization of the fluid/gravity correspondence {exists} for any
non-rotating stationary isolated black-hole horizon. We {have also
found} that such correspondence can be generalized to some
non-stationary black holes. As discussed in {the section of}
introduction, the Petrov-like boundary condition help us to reduce
the degrees of freedom {of gravity to that} of fluid. Such condition
also help us to deal with the compact horizon section cases {without
the use of} {the method of long wavelength expansion}. From the
viewpoint of initial-boundary value problem of Einstein equations,
the Petrov-like boundary condition is also a quite natural choice.
Based on the previous work \cite{IBVP}, $C_{lilj}$ is just the free
boundary data of Einstein equations. This observation {should be
helpful for }us to generalize the Petrov-like boundary condition to
non-vacuum cases {such as in Ref.\cite{ZLNTW_MG}, where} some
generalization for the Einstein-Maxwell case has been discussed.

As discussed in Ref.\cite{ZLNTW_MG}, when we talk the Brown-York
tensor, there should be a counter term which depends only on the
induced geometry of the boundary\cite{BY-counterterm,S-counterterm}
in addition to the extrinsic curvature terms. In our approach, the
Navier-Stokes equation emerges from the conservation law on the
boundary only at the perturbative level, while our boundary
condition requires that the induced geometry should be
fixed\cite{LS-PE2NS}. Therefore, the counterterm has no effect on
the perturbations of the dual fluid on the boundary {and} we can
ignore this term in the procedure.

In previous works \cite{HLPTW_PE2NS-cosmol,ZLNTW_MG}, there is an
artificial coefficient $\alpha$ in the near horizon limit
$r_{c}=\alpha^{2}\lambda^{2}$ as $\lambda\to 0$. The value of
$\alpha$ has been chosen artificially in order to get the
Navier-Stocks equation. In this paper, we have clarified that this
coefficient is related to the surface gravity $\hvar$ of the black
hole {as} $\alpha^2={2\hvar}$.

Because of Eq.(\ref{eq:Cdecomp}), the cases we considered is
space-time without {a} cosmological constant. A little longer but
quite similar calculation can show that the same result also holds
for non-rotating weakly isolated horizon with  {a} cosmological
constant. The cosmological constant gives no contribution for the
gravity/fluid correspondence. This also agrees with the previous
work \cite{HLPTW_PE2NS-cosmol}.

Non-vacuum cases, especially Einstein-Maxwell cases {as well as }
Einstein-Maxwell-Scalar cases, are very interesting topic that will
be considered in future works. Recently, such {setups} have been
used to consider {holographic} magnetofluid\cite{ZLNTW_MG} and
{holographic} superfluid\cite{Liu12}.

The non-rotating condition is crucial for our work. If we give up
this condition, the dual equation will become very complex. For
{instance}, the incompressible condition will lost. How to
understand the gravity/fluid correspondence for rotating black holes
is also an interesting open problem.

Another interesting problem is about the higher order gravity
perturbation. In ref.\cite{CMST11}, Comp\'ere et. al. found , in
Minkowski case, there was a systematic way to construct {such}
correspondence to any order gravity perturbation and to extract the
physical properties of the dual fluid. With the help of long
wavelength expansion, they found the dual fluid system satisfies a
modified Navier-Stocks equations. This dual system is no longer
incompressible and more complex. Interactions between shear, twist
and expansion of fluid will appear in the modified Navier-Stocks
equations. Some recent work\cite{CLYZ12} has {considered} the
Petrov-like boundary condition for higher order perturbation in
Minkowski case. The higher order LS's realization of the
fluid/gravity correspondence {in general spacetime} will be
considered in future work.

\section*{Acknowledgement}
This work is supported by the Natural Science Foundation of China
under Grant Nos.11175245, 11075206,{11275208,11178002. Y.Ling also
acknowledges the support from Jiangxi young scientists (JingGang
Star) program and 555 talent project of Jiangxi Province.}

\end{document}